\documentclass[fleqn,usenatbib]{mnras}

\usepackage{newtxtext,newtxmath}
\usepackage[T1]{fontenc}
\usepackage{ae,aecompl}

\usepackage{graphicx}	% Including figure files
\usepackage{amsmath}	% Advanced maths commands
\usepackage{bm}
\usepackage[usenames]{color}
\usepackage{animate}
\usepackage{mathrsfs}

\graphicspath{{Figures/}}
\usepackage[normalem]{ulem}
\usepackage{array}

\title[Lens dimensionality]{Measuring lens dimensionality in extreme scattering events through wave optics}

\author[Jow et al.]{
Dylan L. Jow,$^{1,2}$\thanks{E-mail: djow@physics.utoronto.ca}
and Ue-Li Pen$^{1,2,3,4,5,6}$
\\
$^{1}$Canadian Institute for Theoretical Astrophysics, University of Toronto, 60 St. George Street, Toronto, ON M5S 3H8, Canada\\
$^{2}$Department of Physics, University of Toronto, 60 St. George Street, Toronto, ON M5S 1A7, Canada\\
$^{3}$Institute of Astronomy and Astrophysics, Academia Sinica, Astronomy-Mathematics Building, No. 1, Section 4,
Roosevelt Road, Taipei 10617, Taiwan \\
$^{4}$Perimeter Institute for Theoretical Physics, 31 Caroline St. North, Waterloo, ON, Canada N2L 2Y5\\
$^{5}$Canadian Institute for Advanced Research, CIFAR program in Gravitation and Cosmology\\
$^{6}$Dunlap Institute for Astronomy \& Astrophysics, University of Toronto, AB 120-50 St. George Street, Toronto, ON M5S 3H4, Canada
}

\date{Accepted XXX. Received YYY; in original form ZZZ}

\pubyear{2021}

\begin{document}
\label{firstpage}
\pagerange{\pageref{firstpage}--\pageref{lastpage}}
\maketitle

% Abstract of the paper
\begin{abstract}
Compact radio sources have been observed to undergo large, frequency dependent changes in intensity due to lensing by structures in the interstellar medium, in so-called ``extreme scattering events" (ESEs). While the study of astrophysical plasma lensing has primarily focused on the geometric limit of optics, coherent radio sources such as pulsars exhibit wave effects when lensed. The additional phase information provided by interference effects in the wave regime may yield more information about the lens than could be obtained in the geometric regime. In this paper, we show that, using wave effects, one can potentially distinguish a one-dimensional lens (where ``one-dimensional" includes both highly elongated lenses, as well as perfectly axisymmetric lenses) from a fully two-dimensional lens, with minimal assumptions on the form of the lensing potential. 

\end{abstract}

% Select between one and six entries from the list of approved keywords.
% Don't make up new ones.

%%%%%%%%%%%%%%%%%%%%%%%%%%%%%%%%%%%%%%%%%%%%%%%%%%

%%%%%%%%%%%%%%%%% BODY OF PAPER %%%%%%%%%%%%%%%%%%

\section{Introduction}
\label{sec:intro}

Extreme scattering events (ESEs), in which radio sources undergo large, frequency dependent changes in intensity, were first observed in quasars \citep{Fiedler1987}. Later, pulsars were also observed to undergo similar changes in intensity \citep{1993Natur.366..320C}. It was realized soon after the first ESE observations that these changes in intensity were not due to intrinsic variation in the source, but rather were due to the refraction of the light by inhomogeneities in the interstellar medium (ISM). Since then, significant effort has gone into understanding the structures in the ISM responsible for these observations.

Most of the physical scenarios proposed to explain the phenomenon of ESEs have fallen into two categories: highly anisotropic lenses and axisymmetric lenses \citep[see e.g.][]{1998ApJ...498L.125W, 1995ApJ...441...70H, 2012MNRAS.421L.132P, dong_extreme_2018}. In both cases, the light is refracted along one direction, and so the lenses can be completely characterized by a one-dimensional lensing potential. Typically, the physical models that produce these potentials propose either spherically symmetric clouds, which when projected onto the lens plane form axisymmetric lenses, or sheet-like structures in the ISM which, under projection, form highly elongated lenses in one dimension. As a result, attempts to model ESE observations have largely focused on highly anisotropic or axisymmetric lenses \citep{1987Natur.328..324R, 1998ApJ...496..253C, 2013A&A...555A..80P, 2016Sci...351..354B, 2018MNRAS.475..867E, 2018Natur.557..522M}. The precise origin and geometry of these ISM structures, however, remains an open question \citep{dong_extreme_2018}. In particular, it remains unclear whether or not fully two-dimensional lenses (i.e. lenses that are neither highly elongated, nor perfectly axisymmetric) may be responsible for some subset of ESE observations.

Thus far, this has been a difficult question to address observationally. Attempts to model ESEs have primarily focused on studying the geometric optics regime where the intensities of the scattered images are added incoherently at the observer. \citet{dong_extreme_2018} argue that axisymmetric lenses with a Gaussian profile can be distinguished from highly anisotropic Gaussian lenses in the geometric optics regime through multiple obersvations of lensing events produced by the same lens, as axisymmetric lenses will tend to produce light curves with more varied morphologies. This relies on the assumption that the lensing potentials in question are all Gaussian. In principle, however, in geometric optics it is always possible to construct a one-dimensional lensing potential that reproduces any given observed light curve, since, in that regime, the intensity is simply equal to the reciprocal of the Jacobian of the lens map. Thus, for any given observation in the geometric regime, it is not possible to determine the dimensionality of the lens absent any prior knowledge of the form of the lens potential allowed. Now, while incoherent sources, such as quasars, are indeed lensed in the geometric optics limit, for coherent sources, such as pulsars, wave optics become important. There has been some attempt to model wave effects in the gravitational lensing literature \citep[see e.g.][]{1986PhRvD..34.1708D, nakamura, 2006JCAP...01..023M, 2013JPhCS.410a2036N, Jow2020}. More recently, there has been a greater effort to understand wave effects in plasma lensing \citep{GrilloCordes2018,2018Natur.557..522M,dongzi2019,Jow2021,2021MNRAS.506.6039S}. Part of the reason for this is the introduction of Picard-Lefschetz theory to the study of lensing, which has helped to ameliorate some of the computational challenges in computing diffraction integrals \citep{job_pl, Jow2021}. 

In this work, we will show that wave effects can be used to directly address the question of whether the lens responsible for a given observation is two-dimensional. In particular, we will show that for one-dimensional lens models, the group delay and the Doppler shift of the images formed near a fold catastrophe follow a predictable relationship. Extra degrees of freedom in the relative motion of the source allow for a violation of this relationship in the two-dimensional case. Observing such a violation would then be strong evidence that a given lens is two dimensional.

\section{One-dimensional lenses}
\label{sec:1D}

A one-dimensional plasma lens lensing a monochromatic, point source of frequency $\omega$ in the thin lens approximation produces a field at the observer that is given by the Kirchoff diffraction integral \citep{Schneider, Nye},
\begin{equation}
    F(\omega, \beta) = \sqrt{\frac{\omega D}{2\pi c i} } \int d\vartheta \exp \Big\{ i \omega \big[ \frac{D}{2 c} (\vartheta -  \beta)^2 - \tilde{\psi} (\vartheta,  \omega) \big] \Big\},
\label{eq:1DF}
\end{equation}
where $\beta$ is the angular position of the source on the sky, and $\tilde{\psi}(\vartheta, \omega) = \Sigma_e( \vartheta) e^2/ (2 m_e c \epsilon_0 \omega^2)$ is the lensing potential. For plasma lenses, the lensing potential depends on $\Sigma_e(\vartheta)$, which is the relative electron surface density in the lens plane as a function of angular position $\vartheta$. When $\Sigma_e > 0$, the lens is over-dense, relative to the surrounding medium, and the lens is divergent. When $\Sigma_e < 0$, the lens is under-dense and convergent. The integral also depends on the distances involved through $D = D_d D_s / D_{ds}$, where $D_d$ is the distance to the lens, $D_s$ is the distance to the source, and $D_{ds}$ is the distance between the source and lens. For cosmological sources, the distances are angular diameter distances and $\omega$ is replaced by $(1+z)\omega$ where $z$ is the redshift of the lens. For purely one-dimensional lenses, the angular coordinates $\beta$ and $\vartheta$ are the projected angles onto the axis of the lens. An axisymmetric lens is, in principle, a 2D lens, but once the angular integral is performed, the diffraction integral is reduced to a 1D integral in the radial direction, where the coordinates $\beta$ and $\vartheta$ are radial angles in the source and lens plane, respectively. Thus, a perfectly axisymmetric lens can be completely reduced to the same formalism as a purely 1D lens. 

It will be convenient to work in terms of dimensionless parameters. Therefore, we can rewrite the phase of the diffraction integral as
\begin{equation}
    i S(x,  y) = i \nu \Big[ (x -  y)^2 + \alpha \psi( x) \Big],
\label{eq:1DS}
\end{equation}
where $x = \vartheta/\vartheta_0$, $y = \beta/\vartheta_0$, for some choice of convenient angular scale $\vartheta_0$, $\nu = \omega D \vartheta_0^2 / 2c$, and $\alpha = - k \Sigma_0 / \omega^2 D \vartheta_0^2$, where $k = e^2/m_e \epsilon_0$, and $\Sigma_0$ is the peak value of the electron surface density in the lens plane. The function $\psi(x)$ gives the shape of the lens potential, and is normalized to a maximum value of unity. Here we assume $\psi(x) > 0$, and define $\alpha$ with a negative sign as we will be primarily considering convergent lenses as examples. Under this definition, $\alpha > 0$ for convergent lenses.  

It is often useful to compute the highly-oscillatory integral in Eq.~\ref{eq:1DF} via the stationary phase approximation, a.k.a. the Eikonal limit. For large frequencies, $\nu \gg 1$, contributions to the integral from isolated points with stationary phase, $\partial_x S = 0$, dominate. The set of points $x_i$ for a given source position, $y$, for which the phase is stationary are given by the solutions to the lens equation:
\begin{equation}
    \xi(x) \equiv y = x + \frac{\alpha}{2} \psi'(x). 
    \label{eq:1Dlenseq}
\end{equation}
The lens equation determines a map, $\xi(x)$, from the lens plane to the source plane, $x \mapsto y$. The diffraction integral is approximated by the sum of the individual contributions of each image, which are given by
\begin{equation}
    F_j = \frac{1}{|\Delta_j|^{1/2}} e^{i S(x_j, y) - i \frac{\pi}{2} n_j},
    \label{eq:1Deikonal}
\end{equation}
where $n_j = -1, 1$ when $x_j$ minimizes or maximizes the phase, respectively. The intensity of the individual images is given by the Jacobian of the lens mapping, which for a one-dimensional lens is $\Delta_j = \xi'(x_j)$. The total wave-field at the observer is then the sum of the individual images given by Eq.~\ref{eq:1Deikonal}. This is distinct from the geometric regime of optics where the total intensity is given by the sum of the individual intensities without the complex phase factor\footnote{Note that some authors refer to the Eikonal limit as geometric optics. Here we use the terms ``Eikonal limit" and ``geometric optics" to distinguish between when the images interfere coherently at the observer and when the intensity of the images are added incoherently, respectively.}. Note that the lens equation, Eq.~\ref{eq:1Dlenseq}, may admit complex solutions, which can have a qualitative impact on the observed field; however, for the purposes of this work, we will restrict our focus to real solutions (see \citet{Jow2021} for a discussion of imaginary images).

The Eikonal limit is extremely convenient to work in, as it allows one to analyse the behaviour of a lens by looking at the individual behaviours of a discrete set of images. For example, each image arrives at the observer with a given relative phase, resulting in an interference pattern at the observer. One can also define a group delay for each image:
\begin{equation}
    \tau_j \equiv \frac{d S}{d \nu} = (x_j-y)^2 - \alpha \psi(x_j),
    \label{eq:1Dgrpdelay}
\end{equation}
where we have used the fact that $\alpha \propto \nu^{-2}$. Thus defined, the group delay $\tau_j$ corresponds to the difference in time it takes for the $j^\mathrm{th}$ image to arrive at the observer relative to the unperturbed image that would exist in the absence of the lens. Note that, in practice, an addition of a constant phase to the diffraction integral (Eq.~\ref{eq:1DF}) is unobservable. Thus, the absolute values of the $\tau_j$ are arbitrary, and it is only the relative group delays between the images that are observable.

Another observable one can define is the Doppler shift of each image (again, relative to some arbitrary reference). A Doppler shift occurs when the source is moving with respect to the lens. For a one-dimensional lens, the Doppler shift arises from motion transverse to the lens, whereas for an axisymmetric lens, the Doppler shift arises from radial motion. In either case, we can write the source position as a function of time, $y(t)$, and define a dimensionless Doppler shift by  
\begin{equation}
    f^D_j \equiv \frac{d S}{d y}. 
    \label{eq:1Ddoppler}
\end{equation}
We have introduced these two observables, $\tau$ and $f^D$, because our goal will be to show that, under certain conditions, there exists a simple relationship between them for images formed by a one-dimensional lenses that is violated for images formed by a two-dimensional lenses. In particular, we will show that a simple relation exists in the case of lensing near a fold catastrophe that is highly off-axis.

Note that this analysis relies on the assumption that the Eikonal limit is valid. That is, our analysis relies on our ability to describe the total flux as a sum of discrete images that are allowed to interfere at the observer. In the low-frequency limit $(\nu \to 0)$, this image description breaks down as all of the images effectively merge into one, in what is sometimes referred to as the ``diffractive" regime of wave optics \citep[see e.g.][]{2005handbook}. Determining the dimensionality of lenses in this regime of wave optics will be more challenging than in the Eikonal regime, and is not the subject of this work.

\subsection{Lensing near a fold}
\label{sec:1Dfold}

In lensing theory, a catastrophe occurs when the lens map is degenerate. The critical curves are the points in the lens plane, $x$, for which the Jacobian of the lens map vanishes. The caustics are the points in the source plane, $y$, that the critical curves are mapped to under the lens mapping. At a caustic, two or more images in the lens plane become degenerate and merge. The degenerate images also attain formally infinite intensities in the Eikonal limit.

\begin{figure}
    \centering
    \includegraphics[width=\columnwidth]{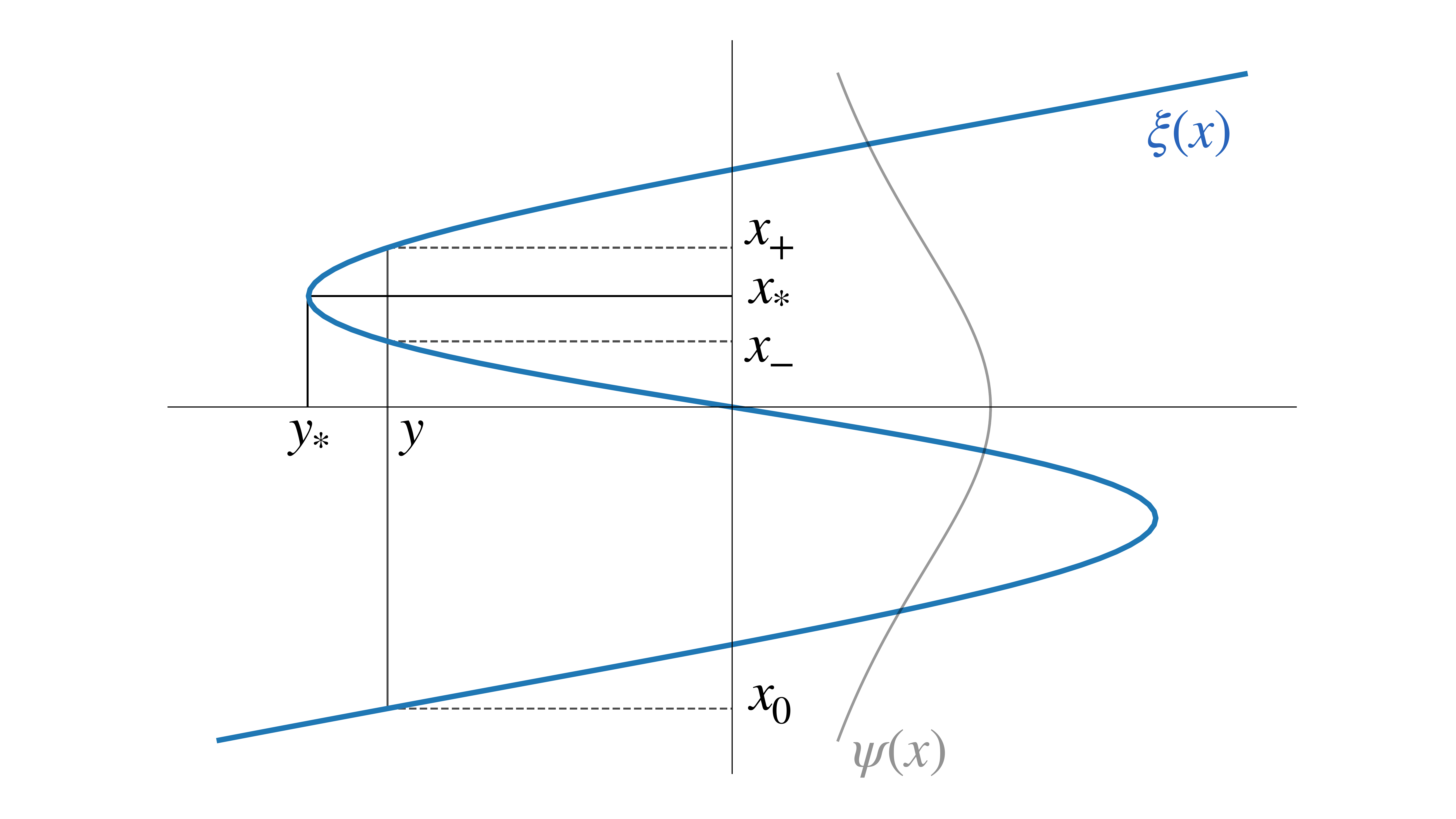}
    \caption{Lens mapping of a generic 1D lensing potential that is symmetric about a central peak at the origin. The vertical axis is the lens plane variable, $x$, and the horizontal axis is the source plane variable, $y$. The blue curve is the lens map $\xi(x)$. The light gray curve shows the shape of the lensing potential, $\psi$, in the lens plane (note that this curve is simply a visual guide, as the height of the lensing potential is not in $y$). The critical point $x_*$ corresponding to one of the folds and its value under the lens map, $y_*$, (the caustic) are shown. For a source position near the caustic, $y$, the corresponding images, $x_0$, $x_-$, and $x_+$ are shown.}
    \label{fig:fold_diagram}
\end{figure}

The mathematical framework of catastrophe theory categorizes the stable catastrophes into a set of characteristic forms \citep[see e.g.][]{Nye}. The simplest form is the fold catastrophe, wherein two images merge. Fig.~\ref{fig:fold_diagram} shows a diagram of the lens mapping for a generic convergent lensing potential with a central peak at the origin that decreases symmetrically away from the peak. Such a potential generically forms two folds. When the source position, $y$, is in between the two caustics, three images are formed, whereas otherwise the lens map is one-to-one and only a single image is formed. 

We will focus on what occurs when the source is near the fold caustic. As shown in Fig.~\ref{fig:fold_diagram}, when the source position $y$ is close to the caustic at $y_*$, three images are formed: two images that are close together, $x_-$ and $x_+$, and another image further away, $x_0$. The two images that are close together are the images that merge at the critical point. The third image, $x_0$, is far from the fold and is effectively un-magnified, and undergoes only minor modulation in intensity compared to the images, $x_-$ and $x_+$, whose intensities diverge at the fold. We will sometimes refer to these two images as the ``perturbed" images, as they undergo large magnifications compared to $x_0$, which we will refer to as the ``unperturbed" image.

Now the group delay of the images given by Eq.~\ref{eq:1Dgrpdelay} can be split up into two parts. The first term is the delay due to the geometric difference in path lengths traversed by each image. The second term is the dispersive delay due to the lens itself. Disregarding the dispersive delay, we can see that the geometric delay of the images must follow a simple relation to the Doppler shift. Namely,  
\begin{equation}
    \tau^\mathrm{geom.}_\pm = (x_\pm - y)^2 = \big( \frac{f^D_\pm}{2\nu} \big)^2,
\end{equation}
where the right-hand side of the equation follows if one assumes that the image locations, $x_\pm$, vary slowly with the source position $y$, i.e. $|\frac{\partial x_\pm}{\partial y} | \ll 1$. Given this assumption, the Doppler shifts are approximately given by
\begin{equation}
f^D_\pm \equiv \frac{dS}{dy} \approx 2\nu(x_\pm - y).
\label{eq:doppler_approx}
\end{equation} 
Thus, when $|\frac{\partial x_\pm}{\partial y} | \ll 1$, the square-root of the geometric group delays of the perturbed images when plotted against the corresponding Doppler shifts form a line through the origin with a slope of \begin{equation}
    m^* \equiv \frac{\sqrt{\tau^{\rm geom.}_+} - \sqrt{\tau^{\rm geom.}_-}}{f^D_+ - f^D_-} = \frac{1}{2 \nu}.
    \label{eq:mstar}
\end{equation}
The inclusion of the dispersive delay somewhat complicates this relationship, but it turns out that when the geometric delay is much larger than the dispersive delay, the contribution to the difference from the dispersive delay has a simple relationship to the contribution from the geometric delay. Namely, we can compute
\begin{align}
    \nonumber
    \sqrt{\tau_+} - \sqrt{\tau_-} &= \sqrt{(x_+ - y)^2 - \alpha \psi(x_+)} - \sqrt{(x_- - y)^2 - \alpha \psi(x_-)} \\
    \nonumber
    &\approx x_+ - x_- - \frac{\alpha}{2} \frac{\psi(x_+)}{x_+ - y} + \frac{\alpha}{2} \frac{\psi(x_-)}{x_- - y} \\
    &= x_+ - x_- + \frac{\psi(x_+)}{\psi'(x_+)} - \frac{\psi(x_-)}{\psi'(x_-)},
\end{align}
where we assume $(x_\pm - y) \gg \alpha \psi(x_\pm)$. In the last line, we utilize the fact that $y = \xi(x_+) = \xi(x_-)$, which is given by Eq.~\ref{eq:1Dlenseq}. 

\begin{figure}
    \centering
    \includegraphics[width=\columnwidth]{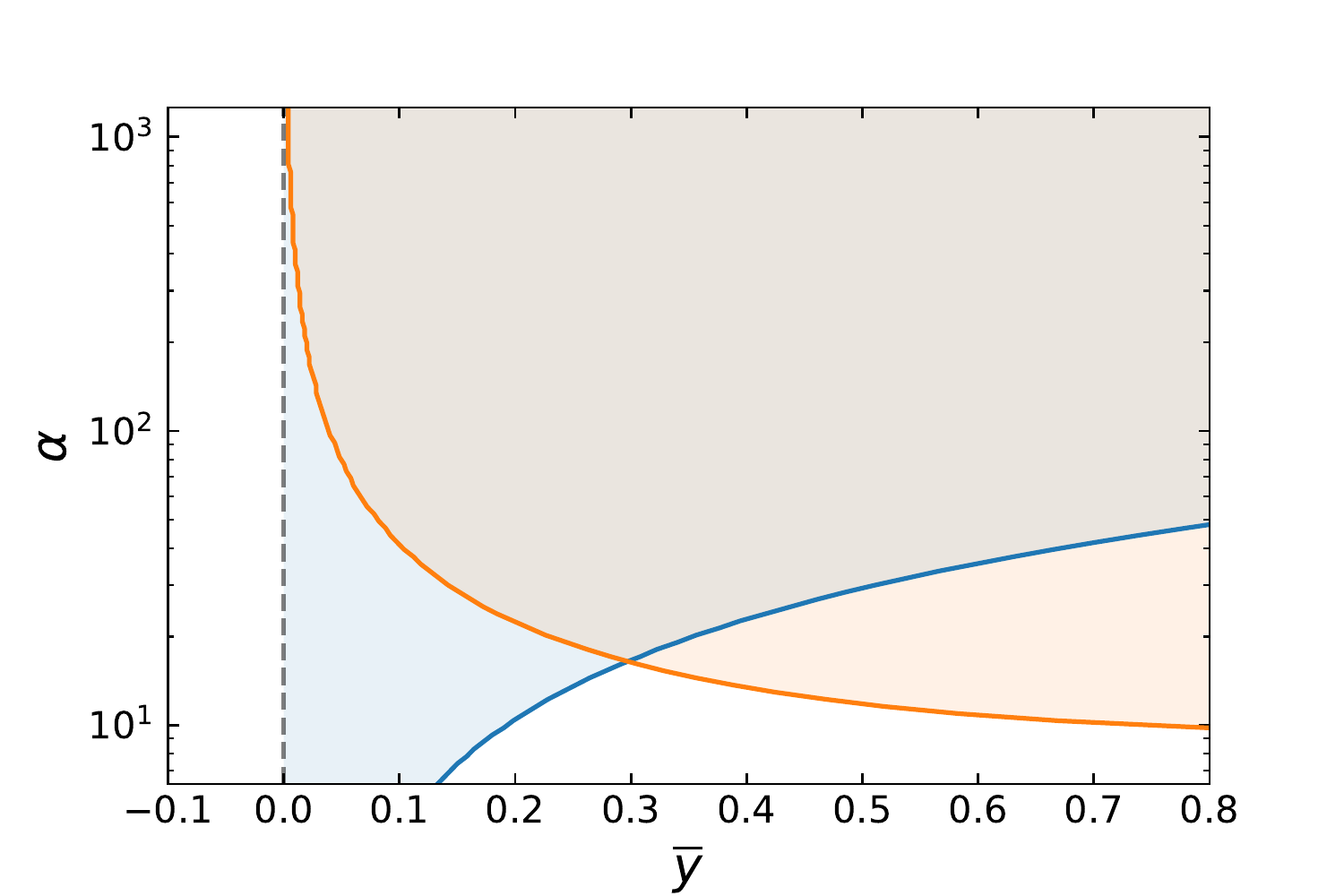}
    \caption{Regions in parameter space for which the assumptions (ii) and (iii) underlying Eq.~\ref{eq:two_slope} hold, for the lens $\psi(x) = 1/(1+x^2)$. The $x$-axis is given by $\overline{y} = y-y*$, where $y^*$ is the location of the left-side caustic (i.e. $y^* < 0$). The blue region corresponds to the region for which $|x_+ - x_-|<0.2$, and the orange region corresponds to the region for which $|\frac{\partial x_\pm}{\partial y}|<0.2$.}
    \label{fig:conditional}
\end{figure}

We can further simplify this expression by recognizing that near the fold $x_+ \approx x_-$. Thus, making the approximation $\psi'(x_+) \approx \psi'(x_-) \approx \psi'(x_*)$, we get
\begin{align}
    \nonumber
    \sqrt{\tau_+} - \sqrt{\tau_-} &\approx x_+ - x_- + \frac{1}{\psi'(x_*)} \big[ \psi(x_+) - \psi(x_-) \big] \\
    \nonumber
    &\approx x_+ - x_- + \frac{1}{\psi'(x_*)} \big[ \psi'(x_*) (x_+ - x_-) \big] \\
    &= 2 (x_+ - x_-).
    \label{eq:sqrt_tau}
\end{align}
Using Eq.~\ref{eq:doppler_approx}, we can relate this to the Doppler shift as
\begin{equation}
    \sqrt{\tau_+} - \sqrt{\tau_-} \approx 2 m^* (f^D_+ - f^D_-).
    \label{eq:two_slope}
\end{equation}
Thus, near the fold, the slope of the two perturbed images in $\sqrt{\tau} / f^D$-space is $2 m^*$. In deriving this simple relationship, we made the following assumptions: 
\begin{enumerate}
    \item the geometric delay is much larger than the dispersive delay,
    \item the source is sufficiently near the fold such that $|x_+ - x_-| \ll 1$,
    \item the right-hand side of Eq.~\ref{eq:doppler_approx} holds, which in turn relies on the assumption that $|\frac{\partial x_\pm}{\partial y}| \ll 1$.
\end{enumerate}
For the generic lens model we have been considering, the first of these assumptions holds when $|y^*|$ is large (i.e. the fold caustic is highly off-axis), which generically occurs when the lens strength $\alpha$ is large. Now, the second and third assumption may appear contradictory at first glance, as it is precisely near the fold that $|\frac{\partial x}{\partial y}|$ becomes large, as $\xi'(x)=0$ near caustics. However, Fig.~\ref{fig:conditional} shows an example of a lens, $\psi(x) = 1/(1+x^2)$, for which these two assumptions can generically hold simultaneously for large $\alpha$. Fig.~\ref{fig:conditional} shows the regions in parameter space for this lens for which  $|\frac{\partial x_\pm}{\partial y}| < 0.2$ and $|x_+ - x_-| < 0.2$, where the $x$-axis is given by the distance to the caustic, $\overline{y} = y - y^*$, and $y^*$ is chosen to be the left caustic as shown in Fig.~\ref{fig:fold_diagram}. As $\alpha$ gets large, the region of overlap for these two conditions gets larger as well. It is only for $\alpha \lesssim 20$ that there is no area of overlap for these two conditions, and the approximation fails.

Note that while Fig.~\ref{fig:conditional} is for a specific example lens, this situation is generic. Near the critical point of a fold, the lens mapping is always quadratic with curvature of order the lens strength regardless of the precise form of the lens potential \citep{Nye}. In other words, the derivative of the lens map is of order $\xi'(x_\pm) = \frac{\partial y}{\partial x} \sim \alpha |x_\pm - x^*| \sim \alpha |x_+ - x_-|$. Thus, both $\frac{\partial x_\pm}{\partial y} = 1/\xi'(x_\pm)$ and $|x_+ - x_-|$ may be less than unity so long as $\alpha \gtrsim 10$. The larger $\alpha$ is, the more likely one may be close to the fold, in the sense that $x_+ \approx x_-$, but far enough away that $\frac{\partial x_\pm}{\partial y} \ll 1$ as the area around the fold where the magnifications exceed unity shrinks as $1/\alpha$.

We will now define a quantity which measures the deviation from the expected relationship given by Eq.~\ref{eq:two_slope}. We define the angle $\delta$ by
\begin{equation}
    \tan \delta = \Big| \frac{m^{\rm obs.} - 2 m^*}{1 + 2 m^* m^{\rm obs.}} \Big|
    \label{eq:delta},
\end{equation}
where $m^\mathrm{obs.}$ is the actual slope of the line formed by the perturbed images in $\sqrt{\tau} / f^D$-space, and $2m^*$ is the expected slope if the lens were one-dimensional. Thus, $\delta$ is the angle between the actual slope and the expected slope given by Eq.~\ref{eq:two_slope}. Note that since our derivation of the expected slope relies on certain assumptions, $\delta$ may still be non-zero for a truly one-dimensional lens, but is, nonetheless, expected to be small. In contrast, however, we will show in Section~\ref{sec:2D} that extra degrees of freedom in the two-dimensional case allow for the maximal value of $\delta = 90^\circ$ to be obtained.

We have already stated that the value $2 m^*$ is an approximate value of the slope when the geometric delay is much larger than the dispersive delay of the two images near the fold. We can examine a specific lensing potential to see when we might expect this condition to be violated for a 1D lens. Consider the lens given by $\psi(x) = 1/(1+x^2)$. The lens mapping produced by such a lens has the same form as shown diagramatically in Fig.~\ref{fig:fold_diagram}. When $\alpha \gg 1$, the fold caustics are located at $y^* \approx \pm \frac{3\sqrt{3}}{16} \alpha$. Thus, the geometric delay of the images near the fold scale like $\tau^{\rm geom.}_\pm \sim y_*^2 \sim \alpha^2$, whereas the dispersive delay scales like $\tau^{\rm disp.}_\pm \sim \alpha$. It follows that the geometric delay dominates the dispersive delay when $\alpha$ is large. When $\alpha=1$, the location of the two fold caustics coincide forming a cusp where all three images merge at a point. Since the cusp occurs at the origin, the geometric delay vanishes, and the dispersive delay is the dominant contribution. 

Eq.~\ref{eq:two_slope} also relies on the assumption that the two images, $x_\pm$, are close together, while the reciprocal of the derivative of the lens mapping at these points is small. As we have already discussed, Fig.~\ref{fig:conditional} shows that for this lens, so long as $\alpha$ is large, one may generically by in a region where both of these conditions are met.

\begin{figure}
    \centering
    \includegraphics[width=\columnwidth]{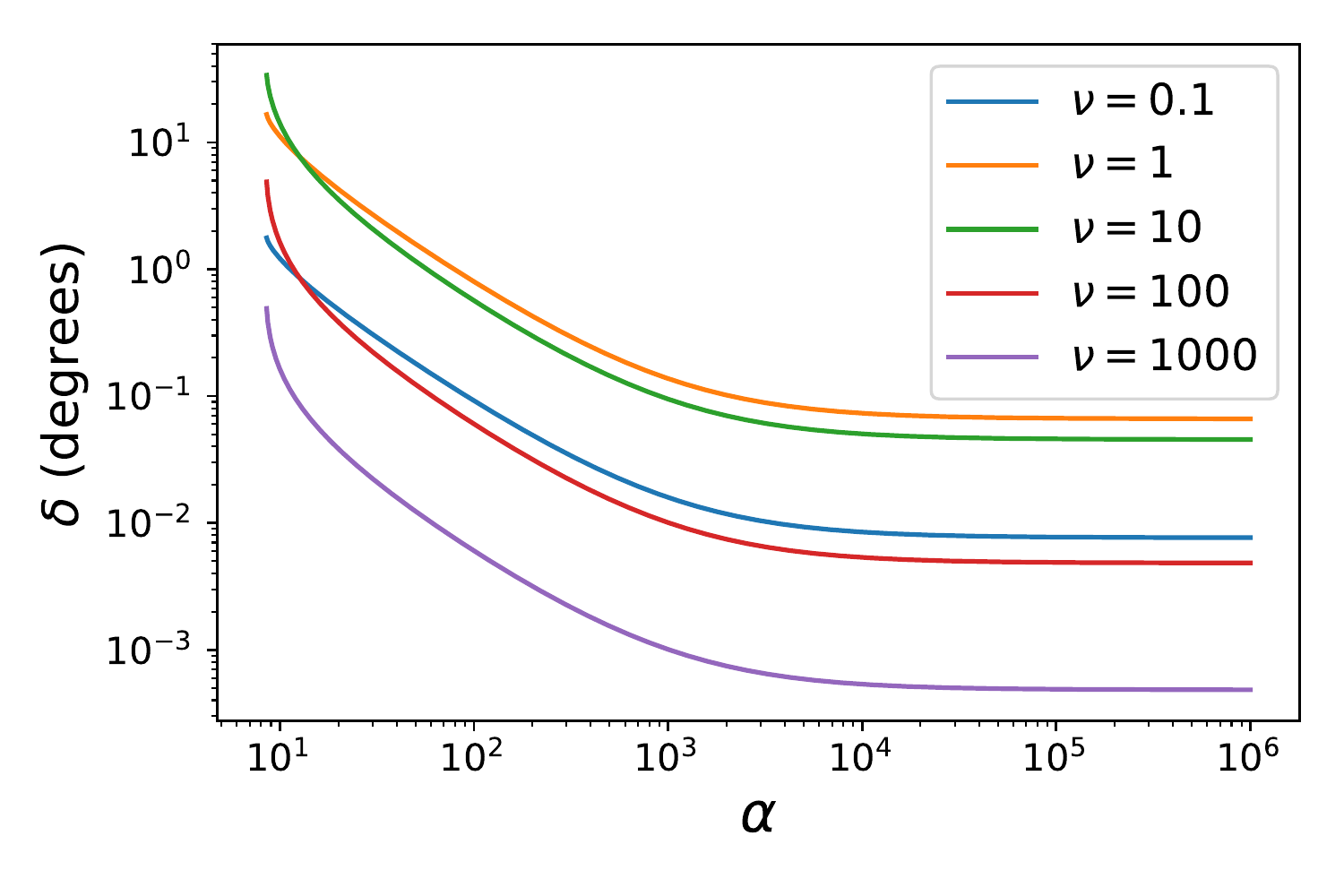}
    \caption{The value of $\delta$ for the 1D rational lens (the angle between the observed slope of the perturbed images near the fold in $\sqrt{\tau}/f^D$-space from the approximate slope of $2m^*$) as a function of $\alpha$, for different values of $\nu$.}
    \label{fig:1D_deltavsalpha}
\end{figure}

Fig.~\ref{fig:1D_deltavsalpha} shows the value of $\delta$ near the fold as a function of $\alpha$ for the rational lens $\psi(x) = 1/(1+x^2)$ for different values of $\nu$ (note that in order to separate the effects of wavelength and lens strength we are treating $\alpha$ and $\nu$ as independent, but it is important to keep in mind that for plasma lenses $\alpha$ is dependent on $\nu$). As one approaches the cusp, i.e. as $\alpha$ decreases, the value of $\delta$ increases, as the dispersive delay becomes comparable to the geometric delay. For large $\alpha$, however, $\delta$ is generally small. Note that this depends on the dimensionless frequency, $\nu$. For frequencies that are either small or large, $m^*$ is either close to zero or infinity, respectively. Thus, deviations of the observed slope from $2 m^*$ lead to only a small difference in angle, $\delta$. For values of $m^*$ close to unity, small deviations in slope are more pronounced as measured by the difference in angle. In any case, when $\alpha$ is large, $\delta$ is small for a 1D lens.

\section{Two-dimensional lenses}
\label{sec:2D}

While one-dimensional lens models are convenient to use for their simplicity, in general, lensing occurs due to a 2D potential on the plane of the sky. Nevertheless, the formalism for 2D lenses is much the same, with the main difference being that the the angular positions in the lens and source planes become vectors. Namely, we make the replacements $x \to {\bm x}$ and $y \to {\bm y}$, and the diffraction integral becomes a 2D integral over the entire lens plane.

The diffraction integral can still be computed using the Eikonal limit, and the locations of the images are given by solutions to the 2D lens equation:
\begin{equation}
    {\bm y} \equiv {\bm \xi}({\bm x}) = {\bm x} - \frac{\alpha}{2} \nabla \psi({\bm x}).
    \label{eq:2Dlenseq}
\end{equation}
The contribution of each image to the Eikonal limit is the same as in Eq.~\ref{eq:1Deikonal}, but the Jacobian is given by
\begin{equation}
    \Delta_j = \det \frac{\partial \xi_k ({\bm x_j})}{\partial {\bm x_l}} 
    \label{eq:2DJac},
\end{equation}
and $n_j = -1, 0, 1$ for when ${\bm x}_j$ is minimum, saddle point, or maximum of the phase, respectively. 

As in the 1D case, catastrophes occur when the Jacobian of the lens mapping is zero. While in the 1D case, the caustics form isolated points, in 2D the caustics form a network of lines in the source plane. These lines are fold caustics and they merge at cusps. As before, we will focus on what occurs near the folds where two images are close to merging. 

Just as in the 1D case, we can define the group delay and Doppler shift of the individual images. The group delay is exactly the same as the 1D case (see Eq.~\ref{eq:1Dgrpdelay}). The Doppler shift, however, is slightly different. Defining $y = |{\bm y}|$, the Doppler shift is given by
\begin{equation}
    f^D_j = \frac{d S}{d y} \approx 2 \nu \big[ ({\bm x_{j,1}} - {\bm y_1}) \hat{{\bm 
    y}}_1 + ({\bm x_{j,2}} - {\bm y_2}) \hat{{\bm 
    y}}_2 \big].
\end{equation}
Note that, again, the right-hand side assumes the image positions are roughly constant with respect to the source position.

We can see more or less immediately from this that the simple relationship derived between the Doppler shift and the group delay in one dimension will not necessarily hold for two dimensions. Whereas in 1D the source and images are all forced to move co-linearly with each other, in 2D there is no such requirement. Indeed, by varying the direction of the velocity, one can cause the slope formed by the perturbed images in $\sqrt{\tau}/f^D$-space to obtain a wide range of values, including reversing the relative order of $f^D_+$ and $f^D_-$.  

\begin{figure}
    \centering
    \includegraphics[width=\columnwidth]{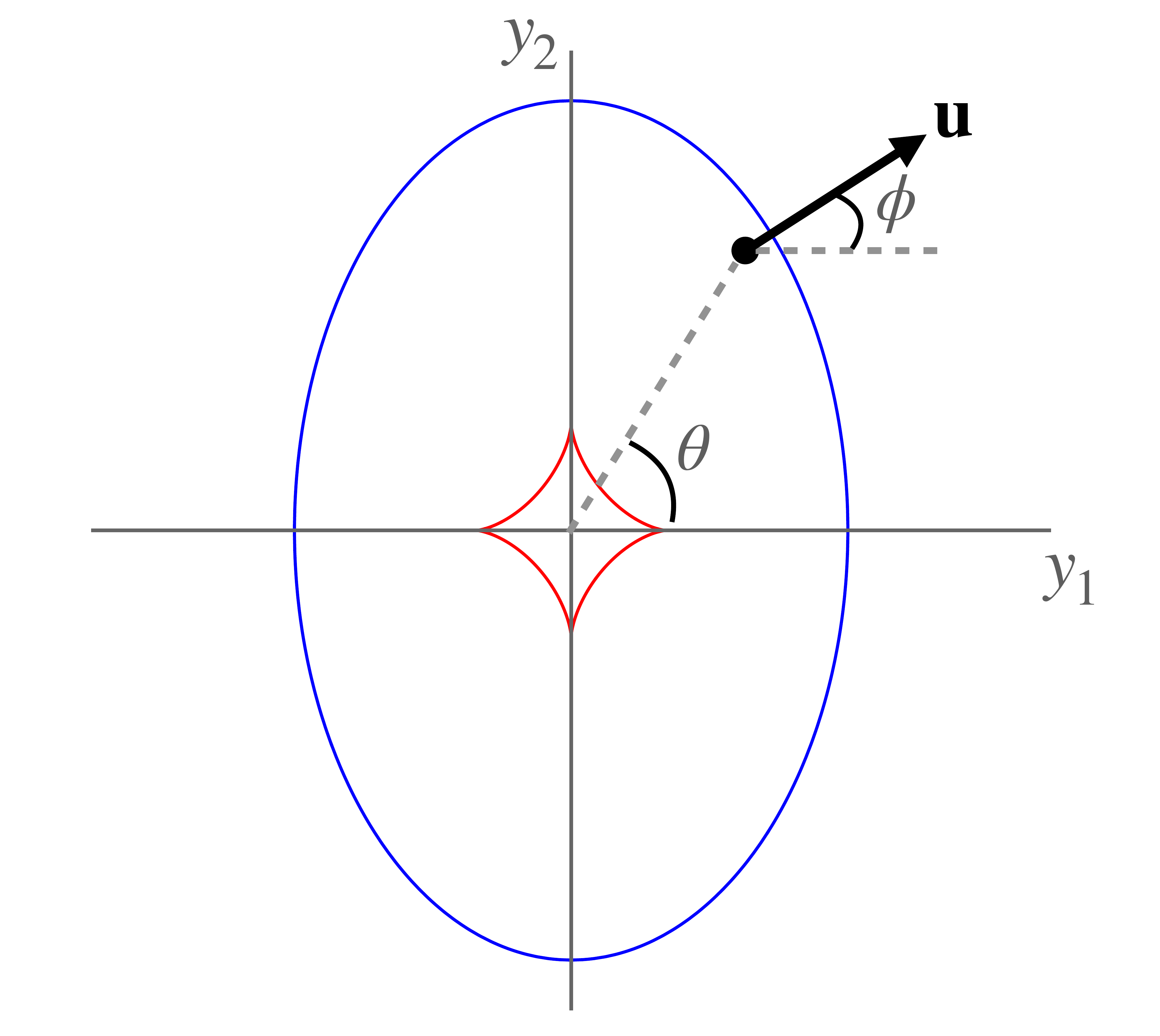}
    \caption{Diagram of the caustic structure of the 2D rational lens given by Eq.~\ref{eq:2D_rat} when $\alpha \gg 1$.}
    \label{fig:2D_caustic_diag}
\end{figure}

As an example, we will consider a 2D rational lens, of the form
\begin{equation}
    \psi_{\rm 2D}({\bm x}) = \frac{1}{1 + {\bm x}^2_1 + 2{\bm x}^2_2}.
    \label{eq:2D_rat}
\end{equation}
Fig.~\ref{fig:2D_caustic_diag} shows a diagram of the caustic structure of this lens when $\alpha \gg 1$. An inner caustic curve (red) is surrounded by a larger caustic curve (blue). The inner caustic encloses a region where the lens equation has five solutions. Between the inner and outer caustics the lens equation has three solutions, and outside the outer caustic there is only one solution. 

\begin{figure}
    \centering
    \includegraphics[width=\columnwidth]{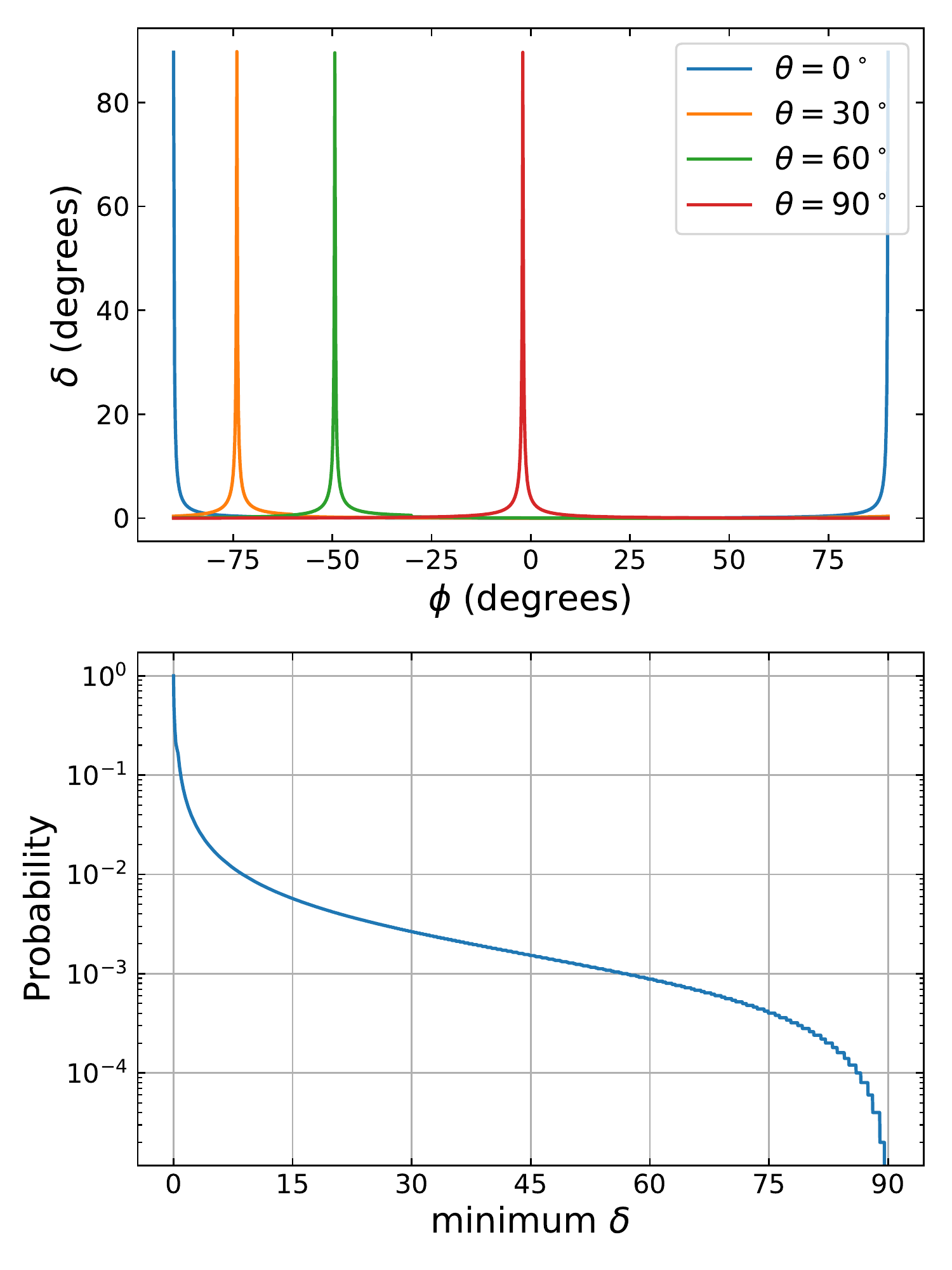}
    \caption{(Top) The value of $\delta$ as defined in Section~\ref{sec:1Dfold} as a function of $\phi$ for the 2D rational lens with $\alpha=200, \nu=1000$, and different fixed values of $\theta$. Fig.~\ref{fig:2D_caustic_diag} shows the angles $\phi$ and $\theta$. (Bottom) The probability that a randomly selected $\phi$ produces a value of $\delta$ greater than some threshold (labeled ``minimum $\delta$" on the $x$-axis) for $\theta=0$ and $\alpha=200, \nu=1000$.}
    \label{fig:2D_deltaphi}
\end{figure}

The top panel of Fig.~\ref{fig:2D_deltaphi} shows the value of $\delta$ as defined in the previous section computed near the outer fold caustic of the 2D rational lens, where the source position for which $\delta$ is computed is given by the angle $\theta$ and the direction of the source velocity is given by $\phi$, with both of these angles shown in Fig.~\ref{fig:2D_caustic_diag}. Since the Doppler shift and time delays depend on both the source position and velocity, as well as the lens parameters $\alpha$ and $\nu$, the angle $\delta$ is a function of all of these parameters, $\delta(\theta, \phi; \nu, \alpha)$. Fig.~\ref{fig:2D_deltaphi} shows $\delta(\phi)$ for fixed $\alpha = 200$ and $\nu = 1000$, and for different fixed values of $\theta$. Fig.~\ref{fig:2D_deltaphi} shows that regardless of the source position, one can always choose a value of $\phi$ that maximizes $\delta$, i.e. that causes the slope of the perturbed images in $\sqrt{\tau}/f^D$-space to be rotated $90^\circ$ from the expected value if the lens were one-dimensional. In fact, the value of $\phi$ at which $\delta=90^\circ$ is given by the direction that is parallel to fold at the source position. 

The bottom panel of Fig.~\ref{fig:2D_deltaphi} shows the probability that, for a fixed source position, $\theta$, the velocity is pointing in a uniform random direction, $\phi$, that results in a value of $\delta$ that is greater than some threshold, assuming the directions are uniformly distributed. In other words, we compute the area in the $\phi$-domain for which $\delta(\phi) > \delta_\mathrm{min.}$ for a fixed value of $\theta$, $\alpha$, and $\nu$. These probabilities are observationally relevant because we want to be able to assess whether a significant deviation from 1D is likely to be observed. If $\delta$ is small, it may not be possible to make a definitive claim that the lens is indeed two-dimensional. 

\begin{figure}
    \centering
    \includegraphics[width=\columnwidth]{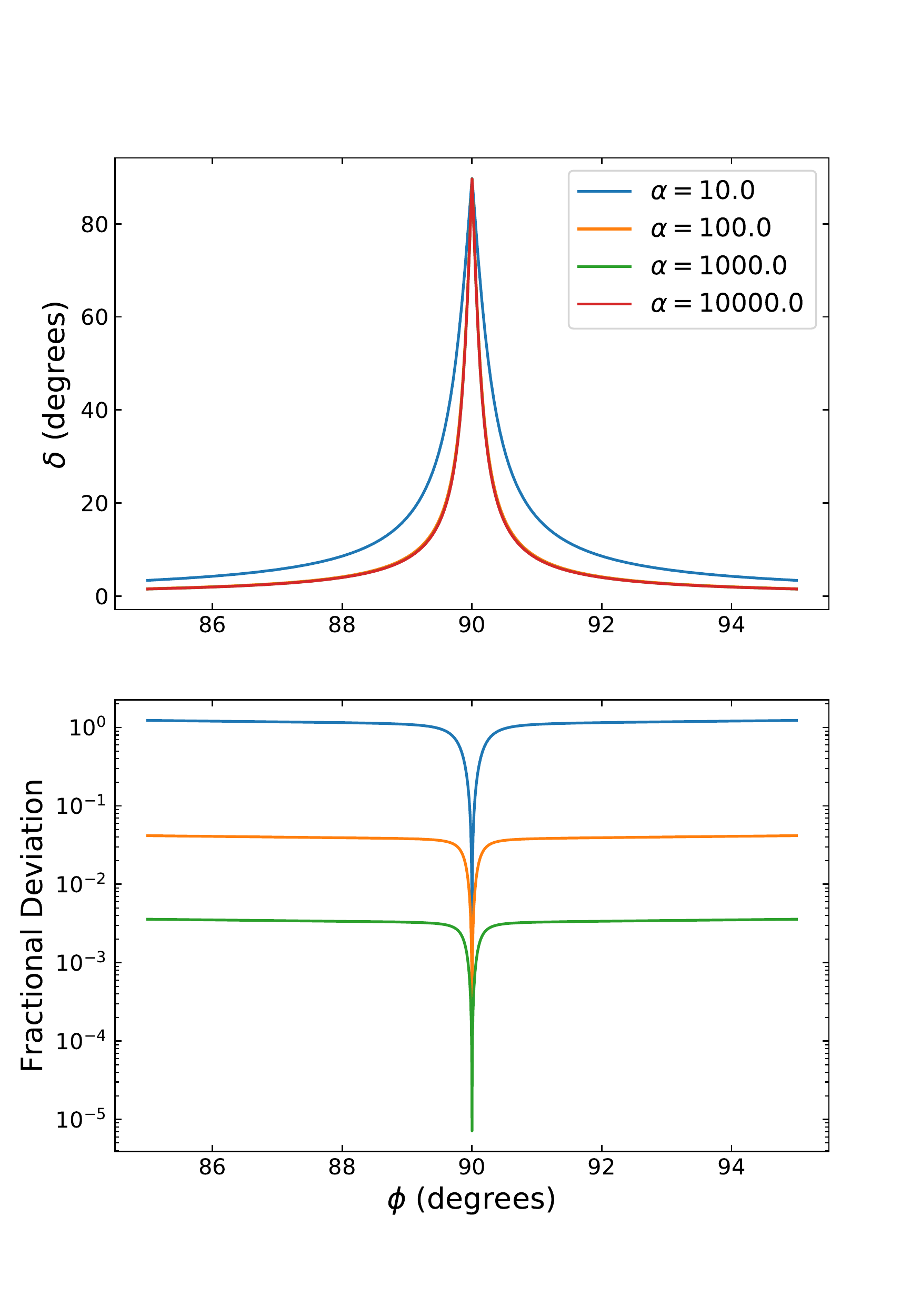}
    \caption{The top panel shows the value of $\delta$ as a function of $\phi$ for the 2D rational lens with $\theta=0$ and different values of $\alpha$. Note that as $\alpha$ increase, by $\alpha = 100$ the curves as function of $\phi$ have all converged. The bottom shows the fractional deviation of the curves in the top panel from the curve for $\alpha = 10,000$.}
    \label{fig:2D_dvsp_alpha}
\end{figure}

Since the effect of the lens is fixed by the parameters $\alpha$ and $\nu$, we want to know how these probabilities vary with them. As in Fig.~\ref{fig:1D_deltavsalpha}, we will treat $\nu$ and $\alpha$ as independent for the following. Fig.~\ref{fig:2D_dvsp_alpha} shows $\delta$ as a function of $\phi$, computed for different values of $\alpha$ for fixed $\theta=0$. As $\alpha$ increases the curves converge. For smaller values of $\alpha$ the width of the curve is slightly larger; however, our focus is on large values of $\alpha$. When $\alpha \approx 3.47$ the outer and inner caustic intersect. As $\alpha$ approaches one, the two sets of caustics merge to a single point: a cusp at the origin. Similarly to the 1D case, at this point the dispersive delay begins to dominate the geometric delay, and the assumptions underlying our analysis cease to hold. 

\begin{figure}
    \centering
    \includegraphics[width=\columnwidth]{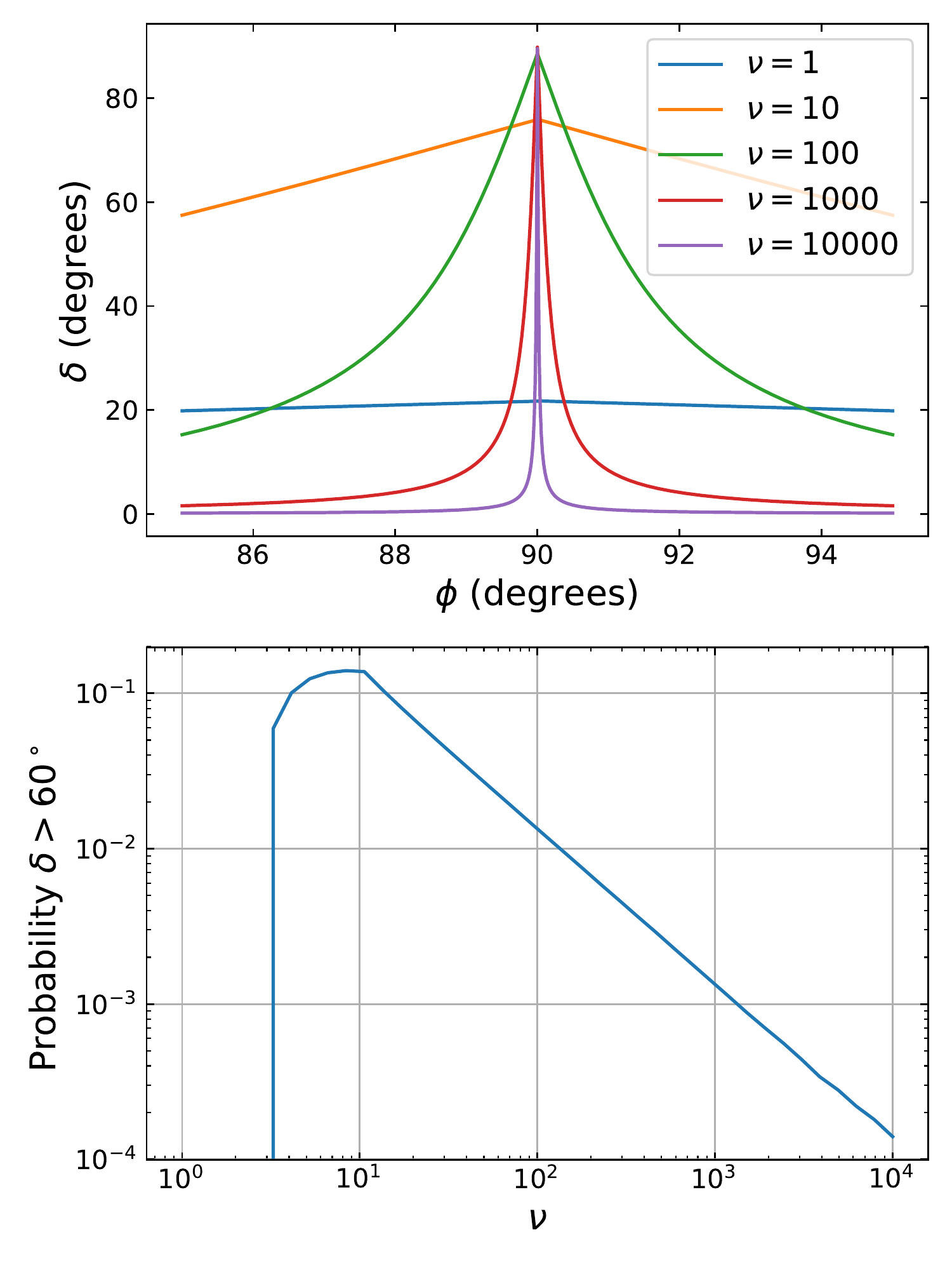}
    \caption{(Top) The value of $\delta$ as a function of $\phi$ for $\theta=0$ and different values of $\nu$ for the 2D rational lens with $\alpha=200$ . (Bottom) The probability that $\delta > 60^\circ$ for uniformly distributed $\phi$ as a function of $\nu$.}
    \label{fig:2D_dvsp_nu}
\end{figure}

The top panel of Fig.~\ref{fig:2D_dvsp_nu} shows the $\delta(\phi)$ curves for different values of $\nu$, for $\theta=0$. The bottom panel shows the probability that a random $\phi$ results in a $\delta > 60^\circ$ as a function of $\nu$. There is a value of $\nu$ for which the probability is maximized. For small $\nu$, the expected slope, $2m^*$, formed by the images near the fold in $\sqrt{\tau}/f^D$-space for a 1D lens is nearly vertical. Thus, to achieve $\delta = 90^\circ$, the observed slope in the 2D case must be flat, but this is not possible due to the finite value of $\sqrt{\tau_+} - \sqrt{\tau_-}$. As $\nu$ gets small, the maximum value of $\delta$ that can be achieved decreases.

So far, we have focused on the case when $\alpha > 0$, i.e. when the lens is convergent, or under-dense. Though there have been physical models proposed for ESEs caused by under-dense lenses \citep[e.g.][]{2012MNRAS.421L.132P}, many of the proposed models feature over-dense lenses \citep[e.g.][]{1998ApJ...498L.125W, dong_extreme_2018}. Our analysis, however, is fairly general, as it only relies on the behaviour of the images near fold catastrophes, which according to Catastrophe Theory, have similar forms regardless of whether the lens is over- or under-dense. We have simply chosen $\alpha > 0$ for concreteness.

\subsection{Measuring $\delta$ from a dynamic wave-field}
\label{sec:measure}

What we have shown in the previous section is that due to extra degrees of freedom in the direction of the source velocity, the simple relationship between the group delay and Doppler shift of the two images near a fold catastrophe that exists in one dimension can be maximally violated in two dimensions. However, it remains to be seen how this effect may be observed in practice from a measurement of the field, $F$, at the observer. 

Generally, one measures the field as a function of frequency and time. The time-dependence of the field arises from the fact that the lens, source, and observer are all moving relative to each other. Thus, one measures a dynamic wave-field, i.e. $F(\omega, t)$. In order to determine the relative group delay and Doppler shift of the images produced by the lens, one can simply perform a 2D Fourier transform along the frequency and time axes. In the scintillometry literature, this is referred to as the conjugate wave-field. Since, in the Eikonal limit, the wave-field is simply the sum of plane waves in the form of Eq.~\ref{eq:1Deikonal}, when taken over a narrow enough bandwidth, the Fourier transform along the frequency axis is approximated by a delta function centred at $\frac{dS}{d\omega}$. Similarly, when taken over a narrow bandwidth along the time axis, the Fourier transform is approximated by a delta function centred at $\frac{dS}{dt}$. Thus, the conjugate wave-field has the form 
\begin{equation}
    \mathscr{F}_{t, \omega} [ F(\omega, t) ] \sim \sum_i \delta(\hat{f}_\omega - \hat{\tau}_i)\delta(\hat{f}_t - \hat{f}^D_i),
    \label{eq:secspec}
\end{equation}
where $\hat{f}_\omega$ and $\hat{f}_t$ are the conjugate variables to $\omega$ and $t$, respectively, and $\hat{\tau}_i$ and $\hat{f}^D_i$ are the dimension-full group delay and Doppler shift of the $i^{\rm th}$ image, given by
\begin{align}
    \hat{\tau}_i &= \frac{dS_i}{d\omega } = \frac{D\vartheta^2_0}{2c} \tau_i, \\
    \hat{f}^D_i &= \frac{dS_i}{dt} = \frac{\mu_\mathrm{rel.}}{\vartheta_0} f^D_i,
\end{align}
where $\mu_\mathrm{rel.} = |{\bm \beta}'(t)|$ is the relative angular velocity between source and lens, which we assume to be constant. For a purely 1D lens, $\mu_\mathrm{rel.}$ is the projected angular velocity along the axis of the lens, and for an axisymmetric lens, $\mu_\mathrm{rel.}$ is the radial angular velocity. 

In order to measure the relative group delays and Doppler shifts of the individual images, in principle one need simply take the Fourier transform along the frequency and time axes of the dynamic wave-field; the group delays and Doppler shifts are given by the peaks in the Fourier transform. The ability to precisely measure these quantities is somewhat complicated, however, by the fact that the phase $S$ is not generally proportional to $\omega$ or $t$. The lens strength, for example, may be frequency dependent (in particular, $\alpha \propto \omega^{-2}$ for plasma lenses). Thus, even in the Eikonal limit, the Fourier transforms of the individual images are not exactly delta functions but have some width, the size of which will be determined by the bandwidth of the variables $\omega$ and $t$. This is further complicated by the fact that for low frequencies, the Eikonal limit ceases to be a good approximation of the diffraction integral. In particular, the field can no longer be expressed as a finite sum of images. The effect of this is to broaden the peaks in the conjugate wave-field for low frequencies, so that it is no longer possible to assign a well-defined Doppler shit and time delay to individual images. Thus, as before, we proceed with out analysis assuming the Eikonal limit holds.

\begin{figure*}
    \centering
    \includegraphics[width=2\columnwidth]{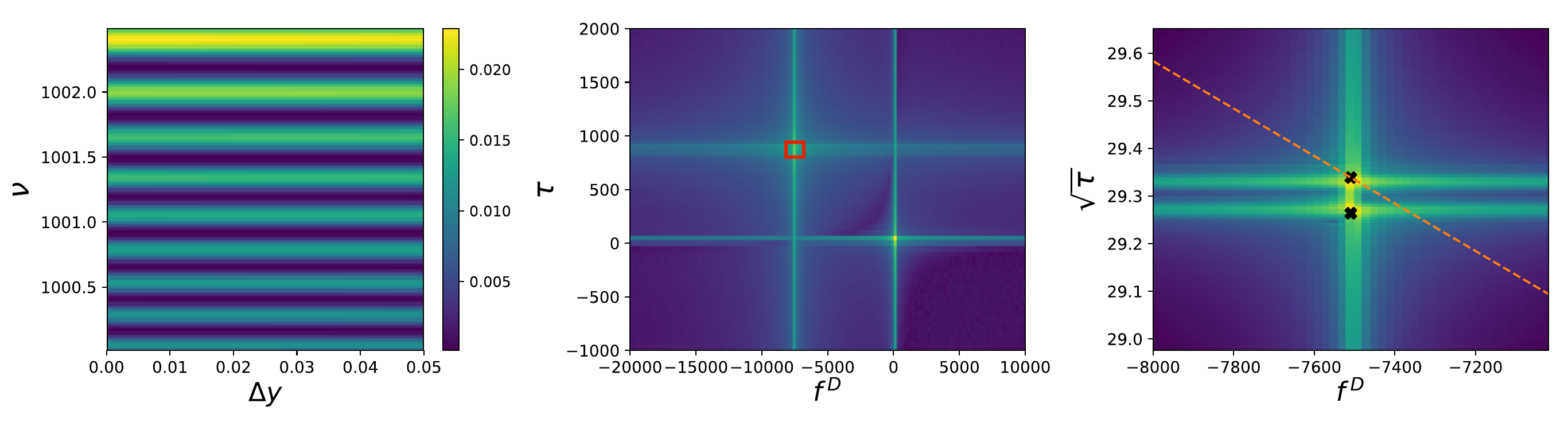}
    \caption{(Left) Amplitude of the perturbed images near the fold of the 2D rational lens (Eq.~\ref{eq:2D_rat}) with $\alpha = 200$. The horizontal axis is defined by ${\bm y} = {\bm y}_0 + \Delta y [\cos\phi, \sin\phi]$, where ${\bm y}_0 = [52.3, 52.3]$ and $\phi = -63.6^\circ$. (Middle) The 2D Fourier transform of the total field along the $\nu-$ and $y-$axes. The variables $\tau$ and $f^D$ are the conjugate variables, and are the dimensionless group delay and Doppler shift. The red box shows the region plotted in the right panel. (Right) A zoomed in version of the middle plot, showing the two perturbed images, with the vertical axis rescaled to be $\sqrt{\nu}$. The orange line corresponds to the expected slope that would have been formed by the two images for a 1D lens.}
    \label{fig:dynspec}
\end{figure*}

Fig.~\ref{fig:dynspec} shows the dynamic and conjugate wave-field of a source being lensed by the two-dimensional rational lens (Eq.~\ref{eq:2D_rat}) with $\alpha = 200$, plotted against the dimensionless quantities $\nu$ and $y$. The horizontal axis of the left panel is defined by ${\bm y} = {\bm y}_0 + \Delta y [\cos\phi, \sin\phi]$, where ${\bm y}_0 = [52.3, 52.3]$ is chosen to be close to the fold, and $\phi = -63.6^\circ$ is chosen to maximize $\delta$ at that point. For these parameters, three images are formed: one unperturbed image far from the fold, and two perturbed images far from the fold (analogous to the one-dimensional situation shown in Fig.~\ref{fig:fold_diagram}. The left panel of Fig.~\ref{fig:dynspec} shows the amplitude of the sum of the two perturbed images near the fold (the amplitude of all three images would be highly oscillatory, and so is not shown). The amplitude of the two perturbed images is roughly constant in $y$ because the $\phi$ that maximizes $\delta$ is parallel to the fold at that point. The amplitude increases as $\nu$ increases, because as $\nu$ increases $\alpha$ decreases ($\alpha \propto \nu^{-2}$), and the fold gets closer to $y_0$. As a function of $\nu$, the amplitude of the perturbed images is given by the Airy function, and the peaks and troughs shown in Fig.~\ref{fig:dynspec} are the Airy fringes that characterize diffraction near a fold. 

The middle panel of Fig.~\ref{fig:dynspec} shows the 2D Fourier transform of the field, $F$, along both the $\nu-$ and $y$-axes. As expected, the Fourier transform takes the form of three narrow peaks in Doppler/delay-space, two close to each other (which correspond to the perturbed images $x_\pm$), and one near the origin ($x_0$). The right panel shows a zoomed in image of the two nearby images, plotted along the square-root of the delay axis. The black $\times$'s show the predicted values of the delay and Doppler shift of the perturbed images in the Eikonal limit. At the resolution computed, the slope formed by these two points is essentially infinite. This is in contrast to what would have been the case for a one-dimensional lens (shown by the orange line), which would have had a slope of $2m^* \sim 10^{-3}$. Thus, for the parameters chosen, the value of $\delta$ is ninety degrees. Note that not only is $\delta$ large for this particular configuration, but for the 2D lens, the ordering of the images in Doppler/delay-space has been reversed. That is, the image with the large magnitude of Doppler shift has a smaller delay. This would not be possible for a one-dimensional lens.

The results shown in Fig.~\ref{fig:dynspec} were computed in the full wave optics regime. The diffraction integral was computed exactly using numerical techniques from Picard-Lefschetz theory. See \citet{2010arXiv1001.2933W}, \citet{job_pl}, and \citet{Jow2021} for a description of how Picard-Lefschetz theory can be used to evaluate highly oscillatory integrals. 

\section{Discussion}
\label{sec:discussion}

Determining the form of the lenses responsible for extreme scattering events is a challenging problem. In particular, there continues to be open discussion regarding whether the lenses responsible are generally highly anisotropic, and therefore can be treated with a one-dimensional lensing formalism, or if a two-dimensional treatment is required.

Many observations of ESEs are for incoherent sources, such as stars, or quasars. Incoherent sources are lensed in the geometric optics limit, where the observed intensity is given by the sum of the individual image intensities, which are determined entirely by the lens mapping, ${\bm \xi}({\bm x})$. In other words, lensing in the geometric optics regime contains no phase information. One consequence of this fact is that a one-dimensional lensing potential can be constructed to fit any given light curve. Thus, one cannot directly determine the dimensionality of a lens in the geometric optics regime.

Coherent sources, however, may exhibit wave effects and should be treated using wave optics. The observed field of coherent point-sources undergoing lensing are given by the Kirchoff diffraction integral. As we have argued, under certain circumstances, the additional phase information in wave optics (encoded by the relative group delay and the Doppler shift of the images) can be used to directly assess the dimensionality of the lens. In particular, we have introduced an observable angle $\delta$ which is small for one-dimensional lenses. Thus, a lensing observation with a large $\delta$ would provide strong evidence that the lens is, in fact, two-dimensional. Note, however, that the converse is not true: a measurement of a small $\delta$ does not necessarily imply that the lens is one-dimensional.

There are some practical issues in making an observation of $\delta$. In principle, one simply needs to observe the field as a function of time and frequency when the source is close to a fold caustic. However, as we noted in Section~\ref{sec:measure}, one's ability to measure group delays and Doppler shifts precisely depends on a variety of observational factors. As we have discussed, our analysis relies on the validity of the Eikonal limit of wave optics. However, when the dimensionless frequency parameter $\nu$ is small, the Eikonal limit ceases to be a good approximation of the diffraction integral. Also, because the lens strength, $\alpha$, is frequency dependent, images only have a well-defined group delay for narrow frequency bands. Similarly, the images only have a well-defined Doppler shift instantaneously. Thus, in order to measure the group delays and Doppler shifts precisely from a conjugate wave-field as described in Section~\ref{sec:measure}, the Fourier transform must be performed over a narrow frequency band and narrow duration. However, as the value of $\nu$ decreases, the Doppler shift increases as well, meaning a longer duration is needed in order to make a measurement of the Doppler shift. Similarly, the bandwidth for which group delays are well-defined decreases for lower values of $\nu$. Thus, it is only in the high $\nu$ limit that one can generally measure these quantities in practice. 

The requirement that $\nu$ be large, however, is in tension with the requirement that $\alpha \gg 1$, as $\alpha \propto \nu^{-2}$. Whether or not the values of these quantities are in the appropriate limit will depend on the parameters of the lens and source in question. Consider, for example, the pulsar PSR B1957+20, which is thought to undergo extreme lensing events due to lensing by the outflow of its companion \citep{2018Natur.557..522M}. Note that these events are not the ``extreme scattering events" we have mentioned previously, as ESEs are due to the ISM. Nevertheless, the lenisng physics is the same, and understanding the lensing of PSR B1957+20, for which there is an abundance of data, may be instructive for the study of ESEs. 

The distance to the pulsar, as well as the distance between the companion and pulsar are known. The typical excess electron surface density of the lenses has also been determined through measurements of the dispersion measure of the lensed pulses. For an observing frequency of $333\,$MHz, the Fresnel scale of this system is given by $R_F = \sqrt{\lambda D_{ds} D_d / D_s} \approx 40\,km$, where $\lambda$ is the wavelength of the source. \citet{2018Natur.557..522M} estimate that the structures in the companion outflow responsible for the lensing events have a size of order $L \sim 1000\,{\rm km}$, and the excess surface density is of order $\Sigma_0 \sim 10^{-3}\,{\rm pc \, cm^{-3}}$. Thus, the dimensionless lensing parameters are $\nu \sim 1800 \, \big(\frac{L}{1000\,{\rm km}}\big)^2$ and $\alpha \sim 20 \big(\frac{L}{1000\,{\rm km}}\big)^{-2} \big( \frac{\Sigma_0}{10^{-3}\,{\rm pc \, cm^{-3}}} \big)$. Thus, the PSR B1957+20 system may be a good candidate for measuring the dimensionality of the lenses in the way we have described. Assuming that the structures in the companion outflow are indeed fully two-dimensional (neither highly anisotropic, nor purely axisymmetric), and taking the values of $\nu \sim 1000$, and $\alpha \sim 40$, we can compute the probability that an observation of any given event for this system can be used to rule out one-dimensional lenses by computing a curve similar to the bottom panel of Fig.~\ref{fig:2D_deltaphi}. The probability of any given event satisfying the requirement that the measured $\delta$ exceed $60^\circ$ is computed to be $\sim 10^{-3}$. Thus, for the system PSR B1957+20, we estimate that roughly 1000 lensing events would need to be observed in order to determine the dimensionality of the lenses responsible.

Ultimately, the point of measuring the group delay and Doppler shifts is to compare the observed relationship between these quantities with the expected relationship for a 1D lens. In principle, this requires \textit{a priori} knowledge of the parameters $D$, $\vartheta_0$, and $\mu_\mathrm{rel.}$, as the expected 1D relationship depends on these parameters. While a precise measurement of $\delta$ would require a precise knowledge of these parameters, it should be noted that a precise measurement of $\delta$ is not necessary to make a strong argument that the lens is two-dimensional. For example, while for 1D lenses, an increase in magnitude of Doppler shift always corresponds to an increase in group delay, this need not be the case for 2D lenses. Observing a decrease in group delay with a corresponding increase in Doppler-shift magnitude would be a smoking gun for a two-dimensional lens, regardless of the values of the underlying parameters. In addition to this, if one measured a large magnitude of the slope formed by the perturbed images in $\sqrt{\tau}/f^D$-space with a corresponding large absolute magnitude of the Doppler values, this would suggest the presence of a 2D lens. While one can, in principle, produce an arbitrarily large slope with a 1D lens by decreasing the value of $\mu_\mathrm{rel.}$ (which is achieved through motion perpendicular to the lens for a highly anisotropic lens, or tangentially for an axisymmetric lens), decreasing $\mu_\mathrm{rel.}$ would also cause a corresponding decrease in the absolute value of the measured Doppler shifts. Thus, for one-dimensional lenses, the slope is only large when the Doppler shift is small.

\section{Conclusion}
\label{sec:conclusion}

We have shown that without any strong assumptions on the form of the lensing potential, $\psi(x)$, one can potentially distinguish between one-dimensional lenses (both highly anisotropic and perfectly axisymmeytric lenses) and fully two-dimensional lenses, through a measurement of the dynamic wave-field near a fold. Since fold crossings are a generic feature of strong lensing events, the method we have described can, in principle, be used to probe the dimensionality of structures responsible for lensing. In particular,  we have identified the binary pulsar system, PSR B1957+20, as a good candidate for such a measurement. Further, the method we have described may also be useful for further probing the structures in the ISM that are responsible for ESEs.

\section*{Acknowledgements}
We thank Fang Xi Lin for useful  discussions. We receive support from Ontario Research Fund—research Excellence Program (ORF-RE), Natural Sciences and Engineering Research Council of Canada (NSERC) [funding reference number RGPIN-2019-067, CRD 523638-18, 555585-20], Canadian Institute for Advanced Research (CIFAR), Canadian Foundation for Innovation (CFI), the National Science Foundation of China (Grants No. 11929301),  Thoth Technology Inc, Alexander von Humboldt Foundation, and the Ministry of Science and Technology(MOST) of Taiwan(110-2112-M-001-071-MY3).
 Cette recherche a \'{e}t\'{e} financ\'{e}e par le Conseil de recherches
en sciences naturelles et en g\'{e}nie du Canada (CRSNG), [num\'{e}ro de
r\'{e}f\'{e}rence 523638-18,555585-20 RGPIN-2019-067].

\section*{Data Availability}
No new data were generated or analysed in support of this research.

\bibliographystyle{mnras_sjf}
\bibliography{biblio} % if your bibtex file is called example.bib

%%%%%%%%%%%%%%%%%%%%%%%%%%%%%%%%%%%%%%%%%%%%%%%%%%

%%%%%%%%%%%%%%%%% APPENDICES %%%%%%%%%%%%%%%%%%%%%

\appendix

%%%%%%%%%%%%%%%%%%%%%%%%%%%%%%%%%%%%%%%%%%%%%%%%%%

% Don't change these lines
\bsp	% typesetting comment
\label{lastpage}
\end{document}